\documentclass[aps,prd,twocolumn,nofootinbib,superscriptaddress]{revtex4}
\usepackage{color}
\usepackage{graphicx}
\usepackage{epsfig}
\newcommand{\be}{\begin{eqnarray}}
\newcommand{\ee}{\end{eqnarray}}

\def\mev{\,\rm MeV}
\def\kev{\,\rm keV}
\def\nue{\nu_e}
\def\anue{\bar{\nu}_e}

\def\lsim{\:\raisebox{-0.5ex}{$\stackrel{\textstyle<}{\sim}$}\:}

\begin{document}




%
\title{Observing supernova neutrino light curve in future dark matter detectors}
\vskip 1cm
\author{Sovan Chakraborty}
\affiliation{Max-Planck-Institut f\"{u}r Physik 
(Werner-Heisenberg-Institut),\ F\"{o}hringer Ring 6,\ 80805 
M\"{u}nchen,\ Germany}
\author{Pijushpani Bhattacharjee}
\affiliation{AstroParticle Physics \& Cosmology (APC) Division, Saha 
Institute of Nuclear Physics,\\
 1/AF Bidhannagar, Kolkata 700064,\ India}
\affiliation{McDonnell Center for the Space Sciences and Department of 
Physics, Washington University in St. Louis, Campus Box 
1105, One Brookings Drive, St. Louis, MO 63130. USA.} 
\author{Kamales Kar}
\affiliation{Ramakrishna Mission Vivekananda University, Belur Math, 
Howrah 711202,\ India}

\begin{abstract}
The possibility of observing supernova (SN) neutrinos through the process of  
coherent elastic neutrino-nucleus scattering (CENNS) in future ton scale detectors 
designed primarily for direct detection of dark matter is investigated. In particular,
we focus on the possibility of distinguishing the various 
phases of the SN neutrino emission. The neutrino emission rates from the recent long term Basel/Darmstadt 
simulations are used to calculate the expected event rates. The recent 
state-of-the-art SN simulations predict closer fluxes among different neutrino
flavors and lower average energies compared to the 
earlier simulation models. We find that our estimated total event rates are 
typically a factor of two lower than those predicted using older simulation models. 
We further find that, with optimistic assumptions on the 
detector's time resolution ($\sim$ 10 ms) and energy threshold ($\sim$ 
0.1 keV), the neutrinos associated with the accretion 
phase of the SN can in principle be demarcated out with, for example, a 
10-ton Xe detector, although distinguishing the neutrinos associated 
with the neutronization burst phase of the explosion would typically 
require several tens of ton detectors. We also comment on the 
possibility of studying the properties of non-electron flavor neutrinos
from the CENNS of SN neutrinos.

\end{abstract}

\maketitle
\section{Introduction}\label{sec:introduction}

The possibilities of studying the propertes of neutrinos and extracting
conditions inside a supernova (SN) progenitor star before and during 
its explosion, through observation of the neutrinos emitted from the 
SN, have been extensively investigated in the past two decades  
following the observation of neutrinos from SN1987A 
\cite{Raffelt:1999tx,duan-09}. Most of these investigations focused on  
neutrino detection using charged current (CC) processes on water and 
other detector materials~\cite{burrows-93,qian-94,choubey-99}, though 
process of inelastic neutral current scattering on, for example, 
oxygen \cite{langanke-96}, deuterium \cite{virtue-01}, and carbon 
\cite{barger-01} were also studied.  

In the last ten years or so, processes like 
neutrino-proton \cite{beacom-02} and neutrino-nucleus \cite{horowitz} 
elastic scattering have also been looked at in detail. In particular, as 
pointed out long 
ago \cite{freedman-74, freedman-77}, there can be the process of 
coherent elastic neutrino-nucleus scattering (CENNS), for which there is 
a significant enhancement in the predicted number of events, increasing 
approximately as the square of the number of neutrons constituting the 
nuclei of the detector material. However, the typical kinetic energies 
of the recoil nucleus in the elastic neutrino-nucleus scattering are 
expected to be rather small, requiring detectors with low energy 
thresholds in the range of few keV to few tens of keV. No conventional 
neutrino detectors have such low thresholds. However, the detectors 
used in dark matter (DM) direct detection (DD) experiments   
designed to look for the weakly interacting massive particle (WIMP) 
candidates of DM through nuclear recoil signals associated 
with WIMP-nucleus scattering may have sufficiently low thresholds so 
as to be sensitive enough to detect the low energy nuclear recoil 
events associated with SN neutrinos \cite{horowitz}. However, the presently 
operating DM detectors are not large enough to detect such neutrino events.
Only in context of planned future large ton scale detectors 
such possibilities would be viable. In the
previous studies \cite{monroe-07,Strigari:2009bq} investigating  
the possibility of observing the CENNS using various astrophysical 
neutrino sources and 
geoneutrinos predict interesting prospects. Indeed, it is 
now well recognized that these CENNS  
events due to astrophysical and terrestrial neutrinos can be a 
major source of background in DM search experiments \cite{monroe-07,Vergados:2008jp,Billard:2013qya}.

\begin{figure*}[!]
\epsfig{file=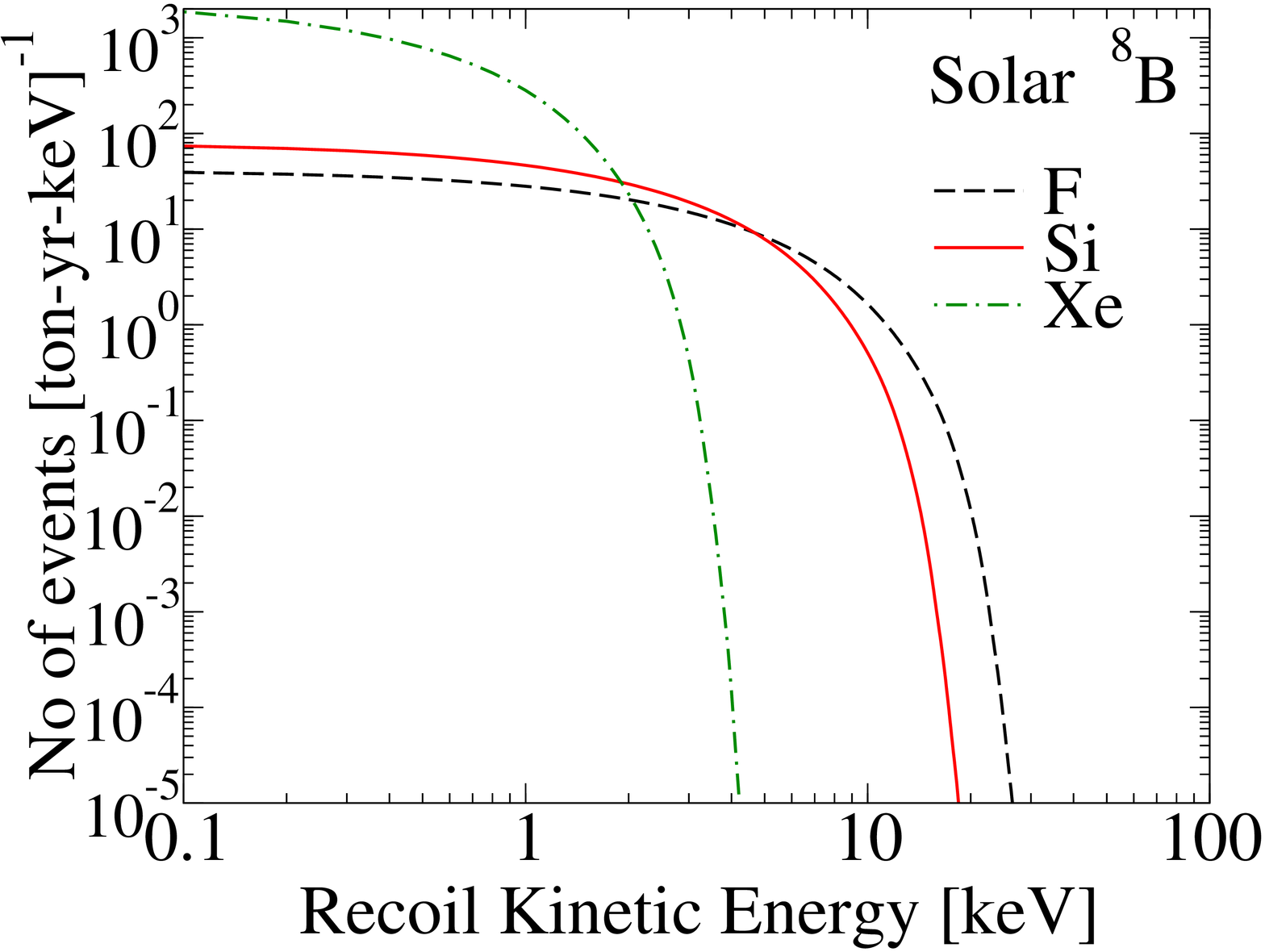,width=0.38\hsize,angle=270}
\epsfig{file=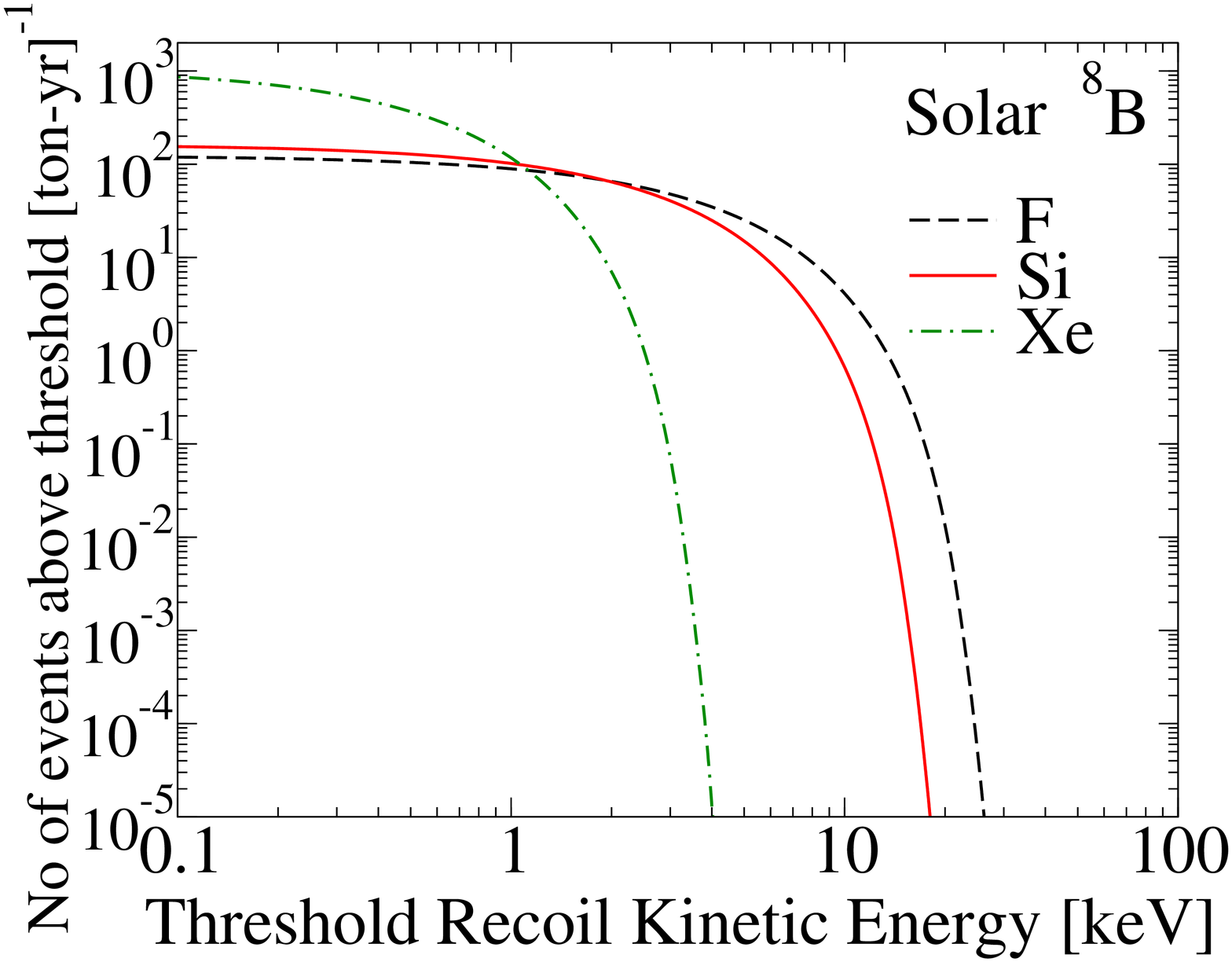,width=0.38\hsize,angle=270}
\caption{Left: Recoil energy spectra (differential event rate as a 
function of recoil nucleus kinetic energy) for $^8$B solar neutrinos in 
a dark matter detector with three different target materials, namely, 
$^{19}$F, $^{28}$Si and $^{131}$Xe. Right: The integral recoil energy 
spectra (total event rate above a threshold recoil energy) as a function 
of the threshold recoil energy of the detector.}
\label{fig:solar_nu}
\end{figure*}
The detectors for DM DD, which are currently in operation 
or are under consideration, use a variety of different target materials. 
In order to optimize the possibility of detecting SN neutrinos using 
such detectors, it is important to study the process of coherent elastic 
scattering of $\sim$ MeV energy neutrinos on various different target 
nuclei. For supernova neutrino detection the DM detectors 
may be particularly useful as the coherent elastic scattering is caused 
by all neutrino/antineutrino flavors as opposed to only the electron 
flavor neutrinos in CC processes 
\cite{horowitz,Scholberg:2012id}. Such coherent scatterings 
will give rise to a few events per $\it{ton}$ of detector material for a 
galactic (10 kpc) SN event 
compared to some hundreds of $\bar\nu_{e}$ events per $\it{kiloton}$  in 
CC based detectors. 
So with an order of magnitude increase in the number of events spread 
over only about 10 
seconds, this needs serious consideration particularly for the future
ton/multi-ton detectors. 

An important consideration for using DM detectors 
for SN neutrino detection is the excellent time resolutions of 
the DM detectors which can be in the region of $\sim$ 10 ms. This offers 
the interesting possibility of studying the temporal structure of the 
neutrino emission from the SN. Indeed, measuring the SN neutrino light 
curve will allow one to probe the standard SN model, which predicts 
three main phases of neutrino emission, namely, the neutronization burst
phase, the accretion phase and the cooling phase. Clear demarcation of 
these phases is extremely important as the flux and average energy of 
the emitted neutrinos are very different in these different phases. 
Since the CENNS process is flavor 
blind, the DM detectors can measure the total SN neutrino light curve, 
and thus will be complementary to the light curves of oscillated $\nue$, 
$\anue$ flavors detected by other detectors through CC interactions. 

In order to derive realistic  estimates of 
the expected number of SN $\nu$ events in a typical DM detector, it is 
important to use a reliable SN model that incorporates as much realistic 
physics of SN explosion and associated neutrino emission as possible. 
In this paper, we use the results from the Basel/Darmstadt 
simulations \cite{Fischer:2009af} to study the detectability of 
SN neutrinos in DM detectors employing various different detector 
materials, and also study the possibility of demarcating the different 
phases of the neutrino emission from the SN using these detectors.  

It may be mentioned here that the average energies of the emitted 
neutrinos given by the Basel/Darmstadt simulations are typically lower
than those given by earlier simulations (see, e.g., \cite{Totani:1997vj,Keil:2002in}). 
Consequently, we find somewhat lower (by about a factor of 2) 
number of events compared to earlier estimates~\cite{horowitz}.
 
This paper is organized as follows: In section 
\ref{sec:Neu-Nuc-coherent}, we briefly review the CENNS process. There, 
for the purpose of illustration, we also 
calculate the expected event rates in DM detectors with different target 
materials, for the case of a guaranteed source of astrophysical 
neutrinos, namely, the $^8B$ solar neutrinos. Section \ref{sec:SNevents} 
contains the main new results of this paper, where we estimate the 
expected number of events from a future Galactic SN in DM detectors with 
different target materials and discuss the possibility of detecting the 
different phases of the neutrino emission. We further  
discuss the possibility of extracting the temperature of the non electron neutrinos 
in the SN neutrino flux. Finally, section \ref{sec:summary} 
summarizes the main results of this paper. 

\section{Coherent elastic neutrino-nucleus scattering}  
\label{sec:Neu-Nuc-coherent}
The differential cross section for the process of coherent elastic  
neutrino-nucleus scattering is given by \cite{freedman-77, Lewin:1996rx}
\begin{equation}
 \frac{d\sigma(E_{\nu},E_{k})}{dE_{k}} = 
\frac{G^{2}_{F}}{4\pi}Q^{2}_{W}MF^2(Q^{2}) 
\left(1-\frac{ME_{k}}{2E^{2}_{\nu}}\right)\,,
\label{eq:coherent_xsec}
\end{equation}
where $E_{\nu}$ is the neutrino energy, $E_k$ is the recoil nucleus 
kinetic energy, $Q_W = N-(1-4\sin^{2}\theta_{W})Z$ is the weak nuclear 
charge for a nucleus with $N$ neutrons and $Z$ protons,  
$M (=AM_N)$ is the mass of the nucleus with mass number 
$A(=N+Z)$ and $M_N=931\mev$ is the average nucleon mass. The form 
factor,  
$F(Q^2)$ (with normalization $F(Q^2=0) = 1$, $Q$ being the 
4-momentum transfer to the nucleus), represents the loss of coherence at 
high recoil energies when $ qR \gtrsim 1 $, $q=(2ME_k)^{1/2}$ being the 
magnitude of the 3-momentum transfer, and $R$ the radius of the 
nucleus. The maximum kinetic energy of the recoil nucleus is $E_{k,{\rm 
max}}=2E_\nu^2/M \approx (2/A)(E_\nu/\mev)^2 \kev$ for $M\gg E_\nu$. In 
the limit of complete coherence ($qR\to 0$), the cross section 
(\ref{eq:coherent_xsec}) varies approximately as the square of the 
number of neutrons in the nucleus. In the following analysis, for 
$F(Q^2)$ we use 
the Helm form factor with the parametrization given in  
\cite{Lewin:1996rx}. 

The differential nuclear recoil event rate is given by 
\begin{equation}
\frac{dN(E_k)}{dE_k} = \int_{E_{\nu,{\rm min}}}^{\infty} 
\frac{dN_{\nu}}{dE_{\nu}}\frac{d\sigma(E_{\nu},E_{k})}{dE_{k}}dE_{\nu}\,, 
\label{eq:event_rate}
\end{equation}
where $\frac{dN_{\nu}}{dE_{\nu}}$ is the differential neutrino flux 
incident on the detector and
$E_{\nu,{\rm min}} = (ME_k/2)^{1/2}$ is the minimum neutrino 
energy required to give a recoil kinetic energy of $E_k$ to the nucleus. 

The CENNS process has not yet 
been observed. To illustrate the kind of event rates one may expect 
in typical DM detectors, with astrophysical sources of neutrinos, we 
display in Figure \ref{fig:solar_nu} the expected event rates as a 
function of the recoil energy (i.e., the recoil energy spectrum) for the 
case of a guaranteed source of astrophysical neutrinos, namely, the 
solar $^{8}B$ neutrinos, and for three 
different target nuclei constituting the detector material, 
namely, $^{19}F$, $^{28}Si$ and $^{131}Xe$. 

The event rates shown in Figure \ref{fig:solar_nu} do not take into 
account the energy resolution of the detector. To illustrate the 
dependence on the threshold recoil energy of the detector, the right 
panel of Figure \ref{fig:solar_nu} shows the integral recoil energy 
spectra above a threshold recoil energy as a function of the recoil 
energy. It is clear that future ton-scale DM detectors have good 
prospects for detecting a neutrino source such as the $^8$B solar 
neutrinos through neutrino-nucleus elastic scattering in a few years of 
running. 
\section{Supernova neutrinos and their detection in dark matter 
detectors}
\label{sec:SNevents}

Detection of neutrinos from SN 1987A \cite{Bionta:1987qt,Alekseev:1987ej,Hirata:1987hu}
established beyond doubt that some supernova explosions are 
associated with emission of large number of neutrinos. According to the 
present understanding of core collapse supernovae, as the core  
of a large star starts to collapse after the nuclear fuel gets 
exhausted, the density in the inner core region at some point goes 
beyond the nuclear matter density. In 
the earlier stages of the collapse, the neutronization
of matter produces $\nu_e$'s, but later at very 
large densities both neutrinos and antineutrinos of all three
flavors ( i.e. $\nu_e$, $\bar\nu_e$, $\nu_{\mu}$, 
$\bar\nu_{\mu}$, $\nu_{\tau}$ and $\bar\nu_{\tau}$) get 
produced in much larger numbers. Almost all the enormous energy 
released in the gravitational contraction, roughly a few
times $10^{53}$ ergs, comes out through the emission of these neutrinos 
over a timescale of $\sim$ 10 seconds.

As already mentioned, the CENNS 
process being flavor blind will be sensitive to $\nu_\mu$, 
$\bar{\nu}_\mu$ and $\nu_\tau$, $\bar{\nu}_\tau$
(hereafter collectively referred to as $\nu_x$) in addition to $\nu_e$ 
and $\bar{\nu}_e$. This will allow a direct estimation of the total 
energy emitted in neutrinos in the SN process. Combined with 
information about $\nu_e$ and/or $\bar{\nu}_e$ event rates for the same 
SN event in other (conventional CC) detectors, this  would then allow 
one to  estimate the average $\nu_x$ energies emitted in the SN event. 

\begin{figure}
\includegraphics[width=.75\columnwidth,angle=270]{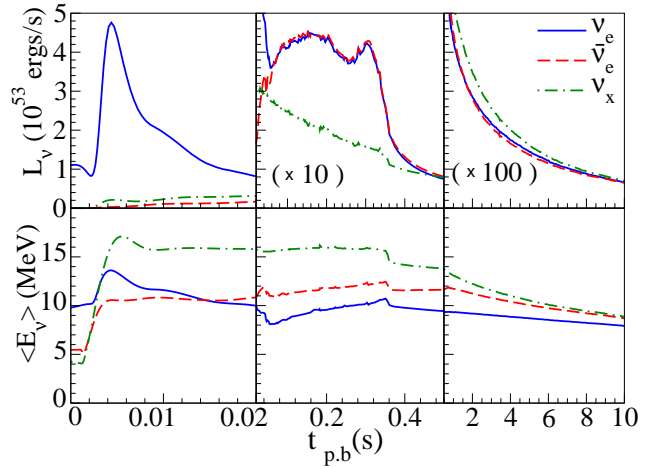}
\caption{Temporal profile of the neutrino luminosity 
(upper three 
panels) and average energy of the neutrinos (lower three panels) for 
different neutrino flavors corresponding to the neutronization phase, 
accretion phase and cooling phase (from left to right, respectively), 
for the Basel/Darmstadt simulation of a 18 $M_\odot$ progenitor SN.
}
\label{fig:lum_av_energy_time_profile}
\end{figure}

While exploring these possibilities one has to be careful, however,  
as the SN neutrino properties like average energy, luminosity and energy 
distribution change with the post-bounce time. There are 
three important stages of neutrino emission in SN: the neutronization 
burst phase, the accretion phase and 
the cooling phase. In Figure \ref{fig:lum_av_energy_time_profile} the 
luminosities and average energies of different
neutrino flavors for the different phases are shown for the 
Basel/Darmstadt simulation. 
\begin{figure*}[!]
\epsfig{file=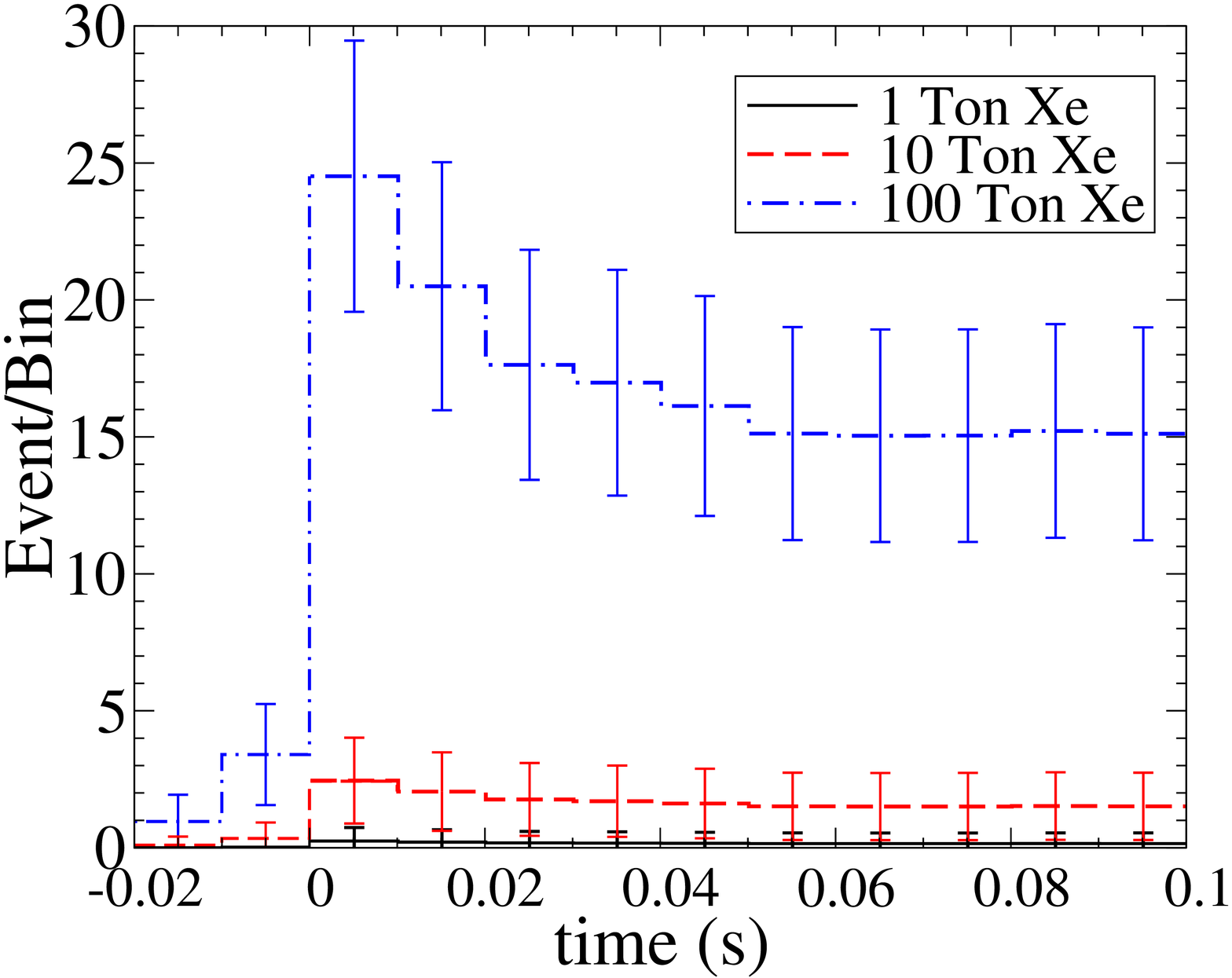,width=0.38\hsize,angle=270}
\epsfig{file=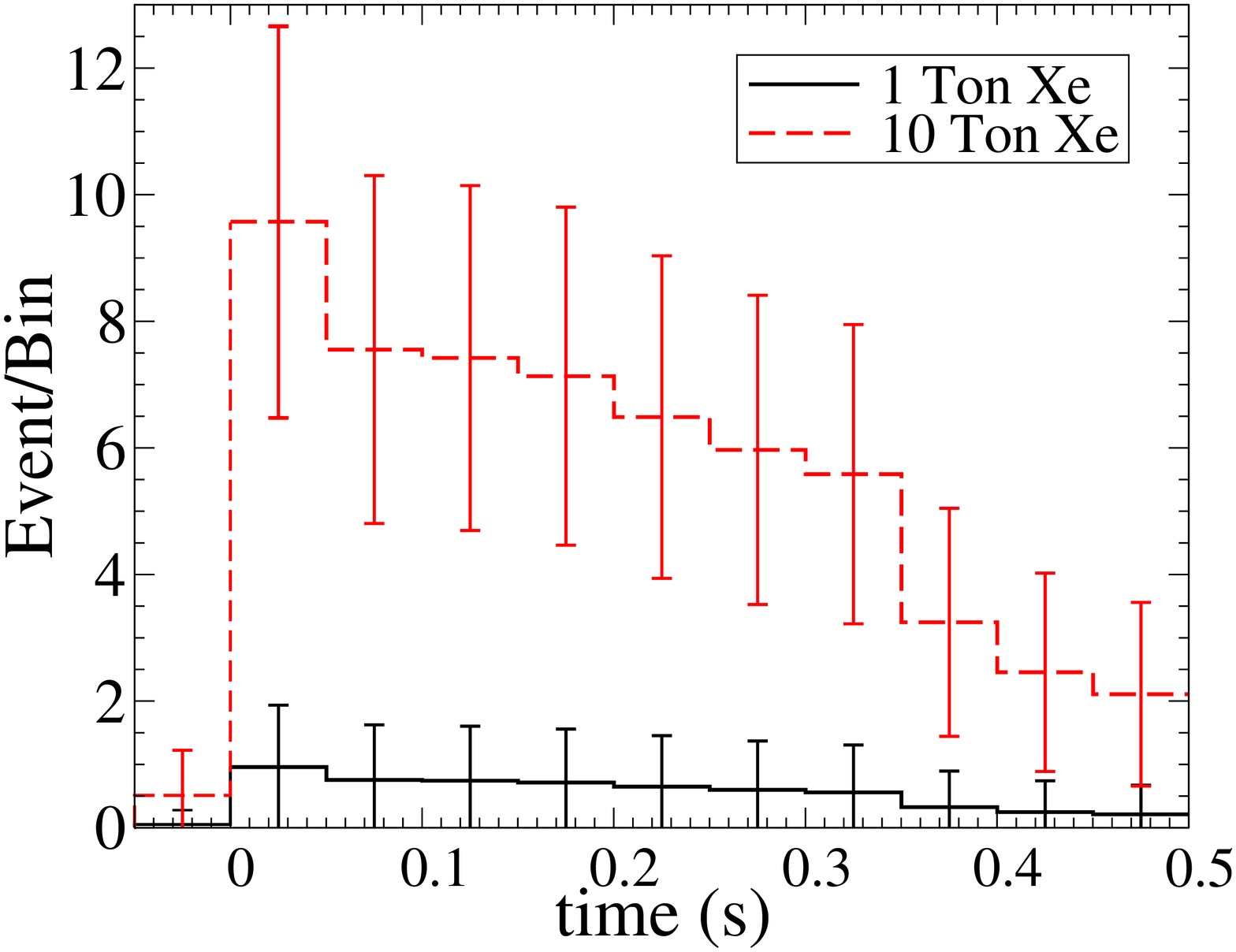,width=0.38\hsize,angle=270}
\caption{Left: Temporal profile of the number of events in 10 ms 
time bins, due to the neutronization burst phase neutrinos from a  
Basel/Darmstadt SN at 10 kpc from earth, in a Xe 
detector with different total target mass assuming a recoil energy 
threshold of 0.1 keV. The statistical 
(Poissonian) errors in each bin are also shown. Right: Same for the 
accretion phase neutrinos in 50 ms time bins. 
} 
\label{fig:nue-burst-accretion}
\end{figure*}
The neutronization burst phase, which is associated with the 
deleptonization of the outer core layers during shock breakout, lasts 
for about 50 ms post-bounce and is characterized by a sharp
peak in the electron neutrino luminosity with little contribution from 
other flavors. This is followed by the accretion phase which is powered 
by infalling matter, and lasts for about 0.5 second. During this 
phase the electron neutrino contribution to luminosity gets 
reduced and the contributions from other species start building up. 
Finally, we have the cooling phase, which lasts for about 10 seconds, 
when all the six species diffuse out of the core. More than 80 percent 
of the total energy emitted in the SN event comes out during this 
cooling phase. The luminosities of all the neutrino species in the 
cooling phase show an approximately exponential decrease with time 
whereas their average energies show a slow linear decrease. 

From the above it is clear that the expected number of events in 
the different emission phases will be different due to 
different luminosities of the emitted neutrinos during these phases. 
Therefore, measuring the temporal structure of the 
detected events will be crucial for extracting the energy spectra of 
the emitted neutrinos. Note also that the flavor oscillation properties 
during the different phases will also be different because of different 
physical conditions during the phases.

\subparagraph{SN neutrino emission spectra:}
 For the initial neutrino distribution in energy and time, we use the 
outputs from the simulation mentioned above \cite{Fischer:2009af}. We 
 factorize the time and energy dependence as 
\begin{equation}
F^0_\nu (t,E_\nu) = \frac{L_\nu (t)}{\langle E_\nu\rangle (t)}\varphi(E_\nu,t)
\end{equation}
for each flavor ($\nu=\nu_e, \overline\nu_e, \nu_x$).
Here $\frac{L_\nu (t)}{\langle E_\nu\rangle (t)}$ represents the 
neutrino emission rate (number of $\nu$'s per unit time) 
with mean neutrino energy $\langle E_\nu\rangle (t)$. The time 
variations of the luminosity and the average energy are taken 
from the simulations. We use the following parametrization of the 
instantaneous normalized ($\int \varphi(E,t)dE = 1$) energy spectrum
$\varphi(E,t) $ from Ref.~\cite{Keil:2002in}: 
\begin{eqnarray}
\varphi(E,t)= \frac{1}{\langle E_\nu\rangle (t)} 
\frac{\left(1+\alpha(t)\right)^{1+\alpha (t)}}{\Gamma 
\left(1+\alpha(t)\right)} 
\left(\frac{E}{\langle E_\nu\rangle (t)}\right)^{\alpha(t)}\, 
\nonumber\\
\times \exp\left[-\left(1+\alpha (t)\right)\frac{E}{\langle
E_\nu\rangle (t)}\right]\, .
 \label{eq:varphi}
\end{eqnarray}
Here $\alpha (t)=\frac{2\langle E_\nu \rangle^2 (t)-\langle 
E_\nu^2\rangle (t)}{\langle E_\nu^2\rangle (t)-
\langle E_\nu\rangle^2 (t)} \,$ 
is the energy-shape parameter~\cite{Keil:2002in} 
and is also extracted from the simulations. Note that all parameter 
values used in the present work correspond to the luminosities and 
average energies of the various neutrino species corresponding to the 
Basel/Darmstadt simulations for a standard 
18$\,M_{\odot}$ progenitor, as shown in Figure 
\ref{fig:lum_av_energy_time_profile}.
 
Having specified the energy and time dependence of the emitted neutrino 
spectra, we now study the possibility of detecting the 
neutrinos from the early emission phases in a typical DM detector. For 
our calculations below we conservatively take the time resolution of the 
detector to be $\sim$ 10 ms. Typical future DM detectors are expected to 
have even better time resolution. So our results presented here should 
be considered as conservative. 

\subparagraph{Neutronization Burst:}
The SN shock while moving outward through the iron core of the SN 
dissociates the iron nuclei, thereby producing free protons and neutrons.
The subsequent electron capture by nuclei and free protons gives rise 
to a large $\nue$ flux, which is emitted in a `burst' when the shock 
breaks out of the neutrinosphere. This deleptonization ``neutronizes'' 
the SN environment. The burst peak shown in Figure  
\ref{fig:lum_av_energy_time_profile} is fairly  independent of the 
details of the SN models such as electron capture rates, nuclear
equation of state and the progenitor mass. In fact, the 
neutronization burst phase is considered as 
the ``standard neutrino candle" for the core-collapse supernovae
scenario, and thus serves as one of the most sensitive probes of the 
physics of neutrino oscillation \cite{Kachelriess:2004ds}
and nonstandard physics \cite{Chakraborty:2012gb}. 

Since the $\nue$ burst phase ends within the first 50 ms, with the time 
resolution of about 10 ms the future DM detectors may have a  
fair possibility of detecting the neutronization burst phase neutrinos.
For illustration, the left panel of Figure \ref{fig:nue-burst-accretion} 
shows the expected number of events in 10 ms time bins in Xe detectors 
of different total target mass for a Basel/Darmstadt SN model (see 
Figure \ref{fig:lum_av_energy_time_profile}) at a distance of 10 kpc 
from the earth. Xe is chosen for this illustration as it offers a 
relatively large $N^2$ enhancement of the CENNS cross section. 
It is clear from Figure 
\ref{fig:nue-burst-accretion} that identification of the neutronization 
burst phase neutrinos will require greater than 10 ton mass Xe 
detectors. For example, a 100-ton Xe detector should be able to 
detect the signature of the neutronization burst phase through neutrinos.   

\begin{figure*}[!]
\epsfig{file=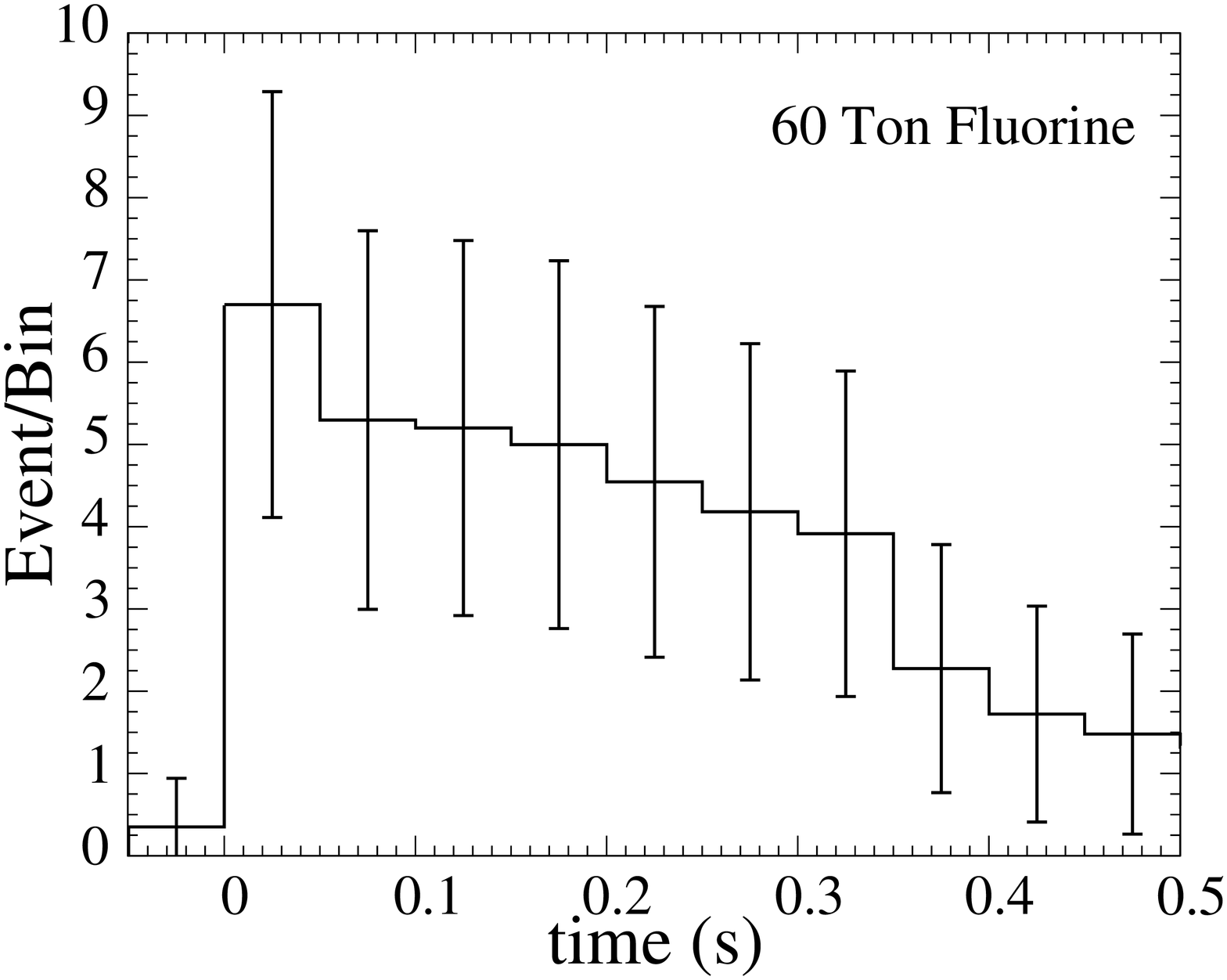,width=0.38\hsize,angle=270}
\epsfig{file=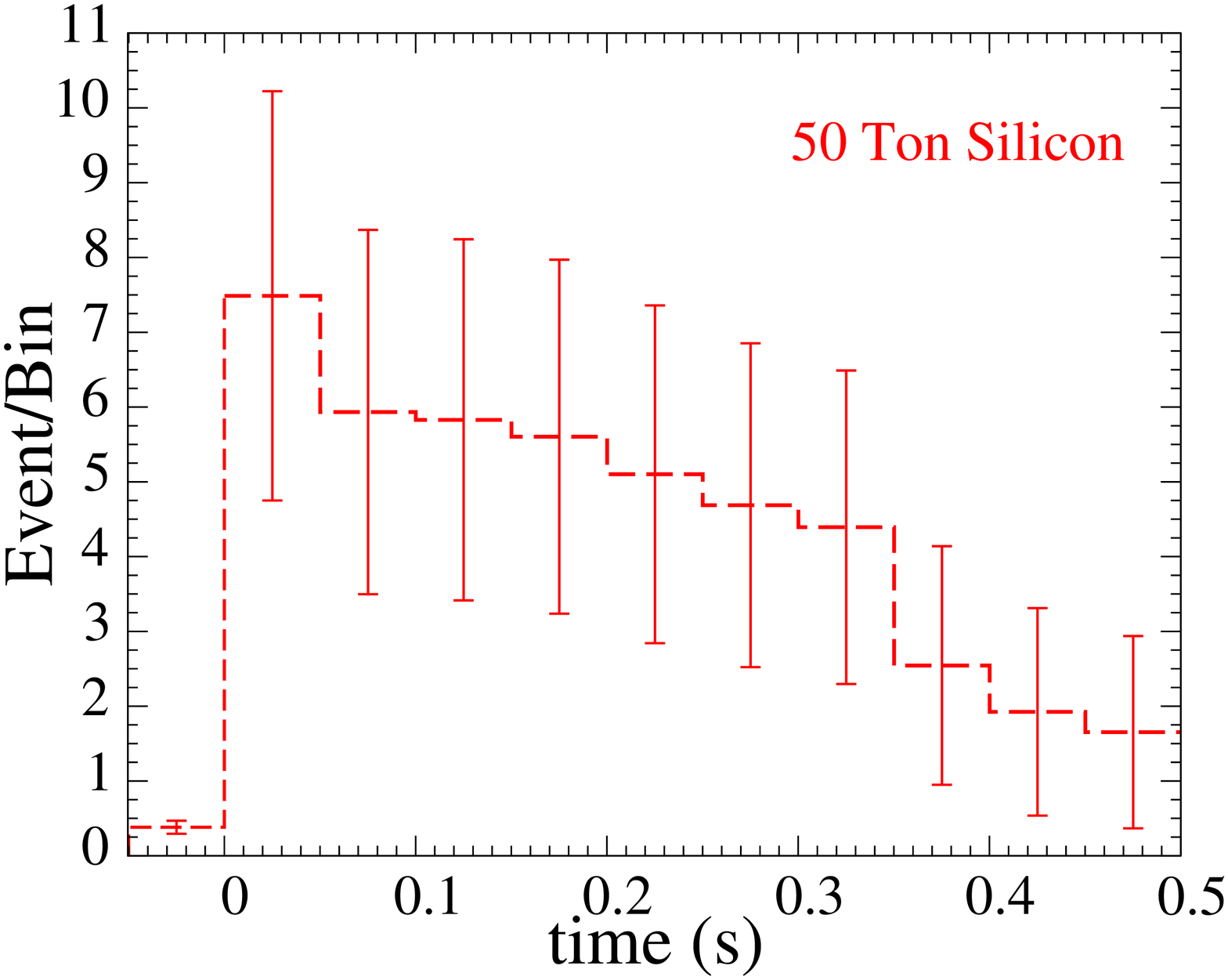,width=0.38\hsize,angle=270}
\caption{Temporal profile of the number of events, due to the accretion 
phase neutrinos from a Basel/Darmstadt SN at 10 kpc from earth, in 50 ms 
time bins in a 60-ton Fluorine (left panel) and a 50-ton Silicon 
(right panel) detector, assuming a recoil energy 
threshold of 0.1 keV. The error bars in both panels are statistical 
(Poissonian).}
\label{fig:SN-F-Si}
\end{figure*}
The estimated numbers of events shown in Figure 
\ref{fig:nue-burst-accretion} assume a zero background and a 
small (0.1 keV) threshold recoil energy of the detector. The backgrounds 
in typical direct DM detectors generally have negligible time variation, 
and the SN events will arrive in a relatively small time window ($\sim$ 
10 seconds), like a pile on the near constant background. Moreover, the 
background from other neutrino sources like the solar and atmospheric 
neutrinos in this small time window would be negligible 
\cite{Arisaka:2012eu} compared to the galactic SN neutrino flux. 
So, the background may not be an issue. However 
the threshold recoil energy of the detector will be important for 
determining the minimum required target mass of the detector for 
detection of the SN neutrinos. Note 
also that the estimated number of events are for a SN at a distance of 10 
kpc; the numbers will scale as the inverse square of the distance to the 
SN. 

\subparagraph{Accretion phase:}
The SN shock loses energy as it moves outward through the iron core, 
finally to stall at around few 100 km from the center of the star. 
At that point, matter starts accreting on to the core, giving increased 
neutrino emission. This process results in the typical accretion hump 
in the neutrino luminosity seen in all SN simulations. The infalling 
material as it accretes onto the core is heated to high 
temperatures thereby allowing electron-positron annihilation resulting
in production of neutrinos of all flavors. Due to high degeneracy of 
$\nue$ and electron during deleptonization, the production of $\anue$ is 
initially suppressed compared to $\nu_x$. However, the $\anue$ flux 
starts growing after the initial deleptonization phase ceases 
as the CC processes (electron and positron captures on free 
nucleons) start becoming more efficient. Interestingly, the 
$\nu_x$, which can be produced only via neutral current processes, can 
not catch up with the $\anue$ and $\nue$. Moreover, being less strongly 
coupled with the environment compared to the electron flavors, 
the $\nu_x$s diffuse out much more swiftly from the SN core than do 
$\nue$ and $\anue$. Hence, $\nu_x$ luminosities remain lower than 
those of $\anue$ and $\nue$. This large 
flux hierarchy between the initial $\nue$, $\anue$ and $\nu_x$ flux 
provides excellent opportunity for studying oscillation physics 
\cite{Serpico:2011ir} with CC based detectors. However, again, the 
CENNS being flavor blind the detectors will measure the total neutrino 
flux and should be able to observe the accretion hump independent of the 
oscillation scenario. Like in the case of the neutronization burst 
$\nue$s, in combination with results from CC based detectors, the 
detection of the accretion hump neutrinos in a DM detector will offer an 
important probe of any new physics and also of the standard SN scenario. 

The right panel of Figure \ref{fig:nue-burst-accretion} shows the 
expected number of events due to accretion phase neutrinos in 50 ms time 
bins in Xe detectors of different total target mass, again for a 
SN at a distance of 10 kpc from the earth. As in the 
case of neutronization burst phase neutrinos, we have assumed zero 
background and a threshold recoil energy of 0.1 keV for the 
detector. Evidently, while a 1-ton Xe detector will not be good enough, 
a 10-ton Xe detector, for example, should be able to pick out the 
temporal profile of the signal. Obviously, a closer SN offers a better 
detection possibility. Clearly, more detailed analysis, including proper 
optimization of the time bin size, threshold recoil energy as well as 
energy resolution of the detector will be required to have accurate 
estimation of the required detector mass.  
\begin{figure*}[!]
\epsfig{file=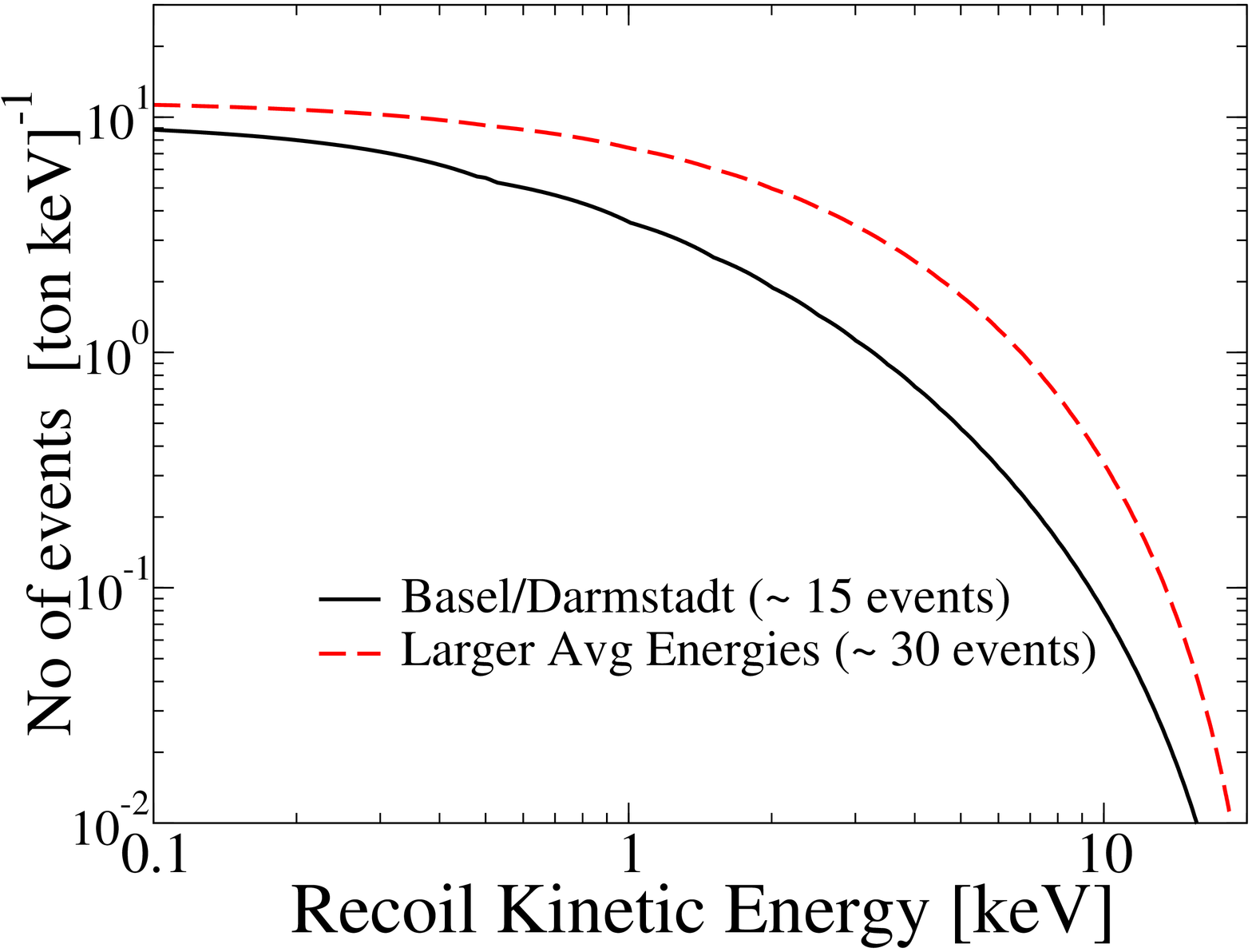,width=0.38\hsize,angle=270}
\epsfig{file=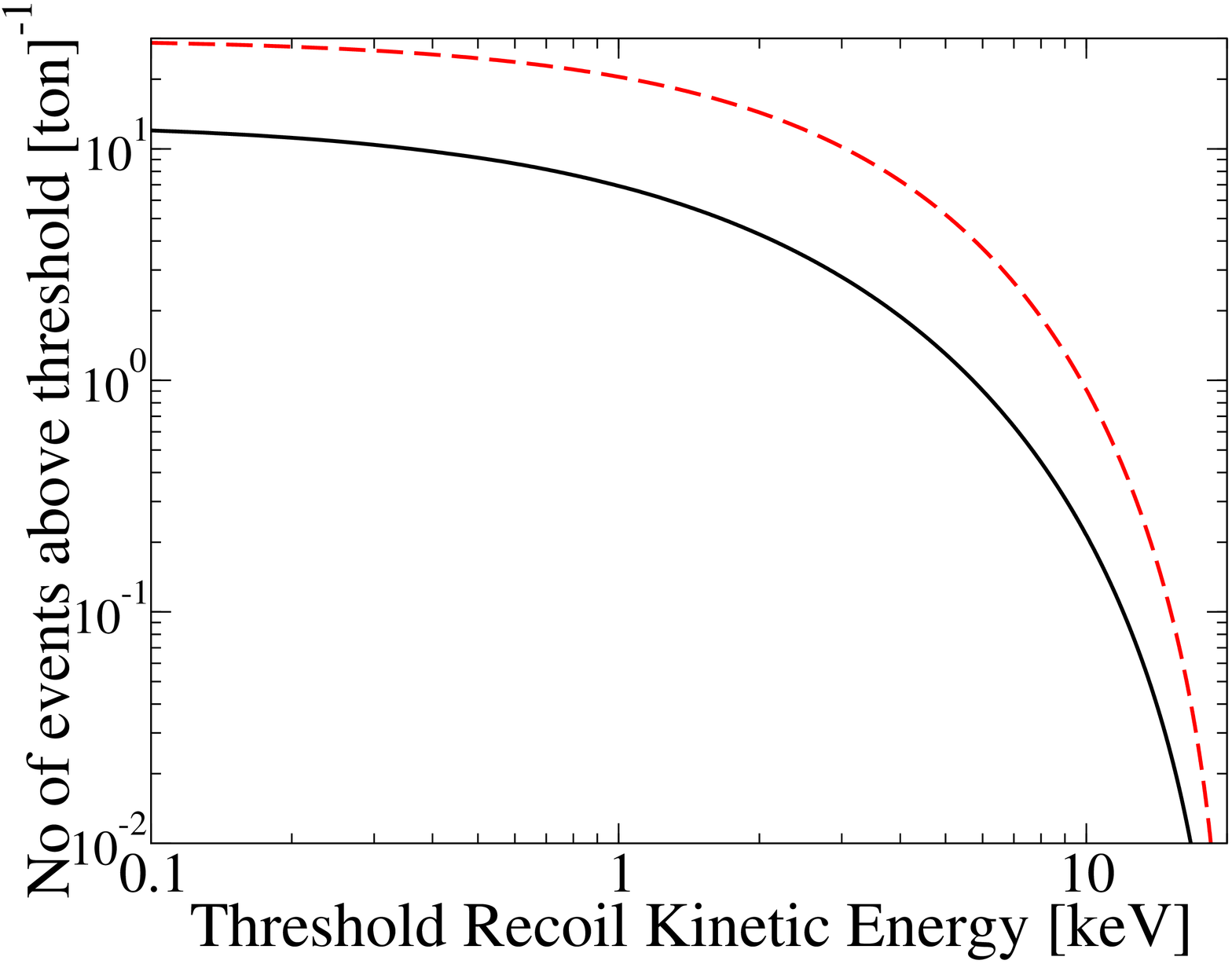,width=0.38\hsize,angle=270}
\caption{Recoil energy differential spectra (left) and integral spectra 
as a function of the threshold recoil energy (right) for SN neutrinos in 
a 1-ton Xe detector. Curves are shown for the Basel/Darmstadt SN model 
as well as for another SN model with average energies of $\nu_e$, 
$\bar{\nu}_e$ and $\nu_x$ equal to 10, 12 and 18 MeV, respectively, 
both for a SN at a distance of 10 kpc from earth. 
}
\label{fig:SN-Xe}
\end{figure*}

\subparagraph{Different Materials:}
Our discussions so far were concerned with Xe as the detector target 
material. However, DM detectors use a variety of other materials. To 
study the suitability of relatively lower mass number nuclei for SN 
$\nu$ detection, below we consider the cases of Fluorine and Silicon, 
which are also used in DM DD experiments. Here, again, we use the same 
Basel/Darmstadt SN neutrino fluxes as described above for a SN at a 
distance of 10 kpc. We optimize the required detector mass so 
that the neutronization burst phases events are are detectable. 
The condition used is that the detector mass should be large enough so
that the lower limits of the predicted number of events including
the statistical fluctuations are above zero for the first five
10-ms time bins during the neutronization burst.
We find that, with this 
criterion for F and Si, one would require at least 60 and 50 ton mass 
detectors, respectively, for detecting the neutronization burst phase 
neutrinos. As expected the required minimum detector 
masses are larger compared to that for Xe (see Figure  
\ref{fig:nue-burst-accretion}). A somewhat better prospect for detection 
is offered by the accretion phase neutrinos, which is shown in Figure 
\ref{fig:SN-F-Si} assuming, again, zero background and a recoil energy 
threshold of 0.1 keV. Notice that the accretion hump is visible 
in both cases. 

\subparagraph{Recoil Energy Spectra:} 
Although demarcation of the temporal profiles (the light curve) of the 
different neutrino emission phases would, as discussed above, require at 
least a 10 ton Xe detector, a smaller, 1 ton detector would be 
good enough to detect a galactic SN event. Figure \ref{fig:SN-Xe} 
shows the expected recoil energy spectra (event rate as a function of 
the recoil energy) in a 1 ton Xe detector for a SN at a distance of 10 
kpc. 

For comparison, in addition to the Basel/Darmstadt SN model 
considered throughout this paper (see Figure 
\ref{fig:lum_av_energy_time_profile}), we also show in 
Figure \ref{fig:SN-Xe} 
the event rate expected for a neutrino flux parametrization with average 
energies of 10, 12 and 18 MeV for $\nue$, $\anue$ and 
$\nu_x$, respectively, for the entire duration of the SN neutrino 
emission as used in previous studies (see, e.g., Ref.~\cite{horowitz}). 
Evidently, the larger $\nu_x$ average energy used in earlier studies  
yielded more expected number of events compared to those for the 
recent Basel/Darmstadt simulations. The relatively larger 
average energies quoted in many previous studies 
represent the early accretion phase of the SN. The average energies in 
the accretion phase are expected to be larger than those in the 
cooling phase. Thus, using the larger average energies for the entire SN 
duration would give larger number of events. We find, for the 
Basel/Darmstadt SN model, the total number of events are reduced by about 
a factor of two compared to earlier estimates \cite{horowitz}. Note, the recent 
results \cite{MartinezPinedo:2012rb} from the same Basel/Darmstadt group shows a larger flux difference 
of different flavors compare to the their old simulation results \cite{Fischer:2009af}. However, 
even the new differences are way too smaller compare to the very large average energy  simulations
of \cite{Totani:1997vj} and thus the overall results remain robust under the new simulations \cite{MartinezPinedo:2012rb} as well.

As clear from the above Figures, the main contribution to the event 
rates comes from recoil energies below 1 keV. Hence only detectors with 
relatively low recoil energy threshold $\lsim 1 \kev$ will have a 
reasonable chance of detecting the SN neutrinos. On the optimistic side, 
however, since the event rates scale as the inverse square of the 
distance to the SN, some of the potential SN candidates, such as 
Betelgeuse, Mira Ceti and Antares, which are relatively close by stars 
(at $< 0.2$ kpc), offer interesting possibilities for the next 
generation DM detectors.     
\begin{figure*}[!]
\epsfig{file=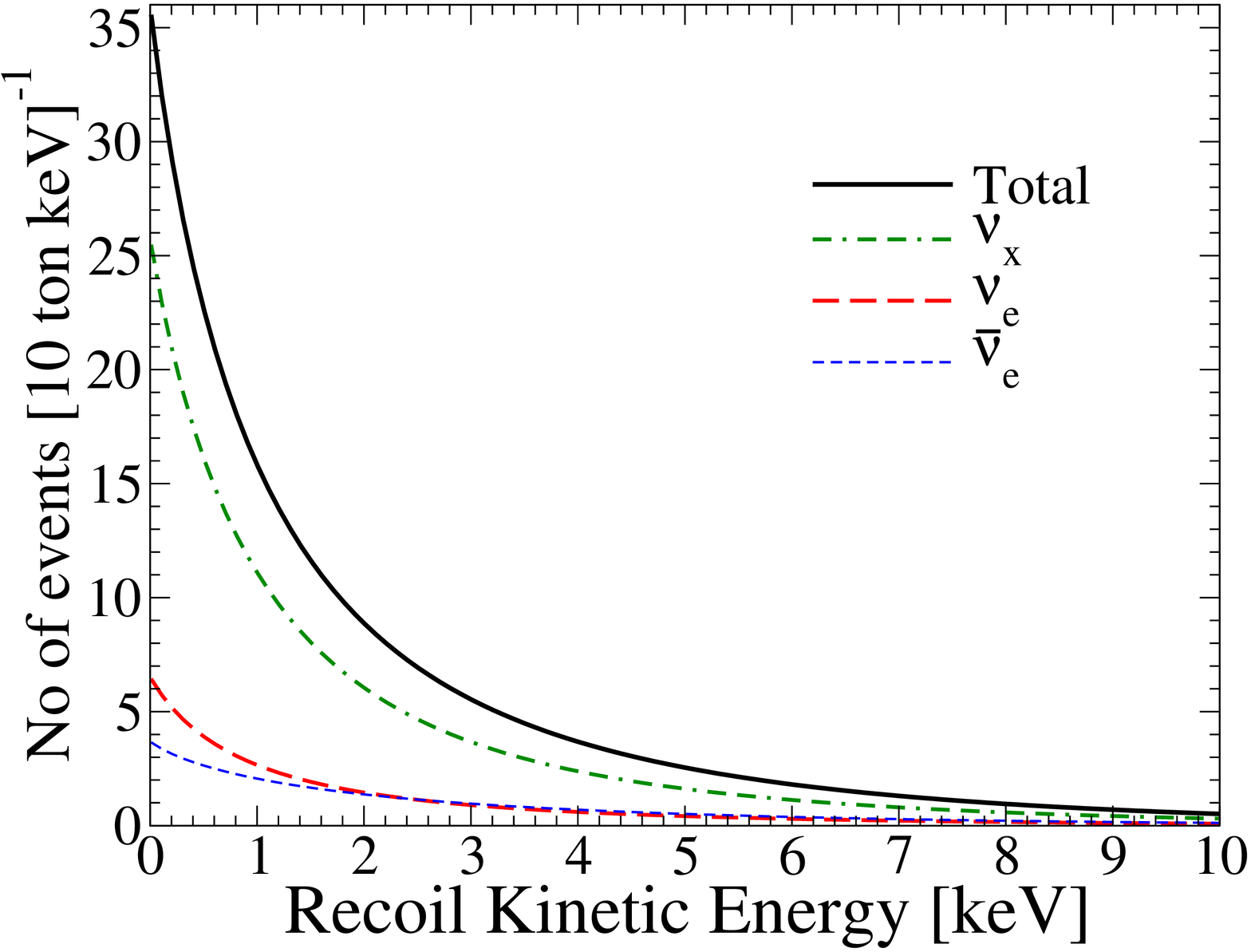,width=0.38\hsize,angle=270}
\epsfig{file=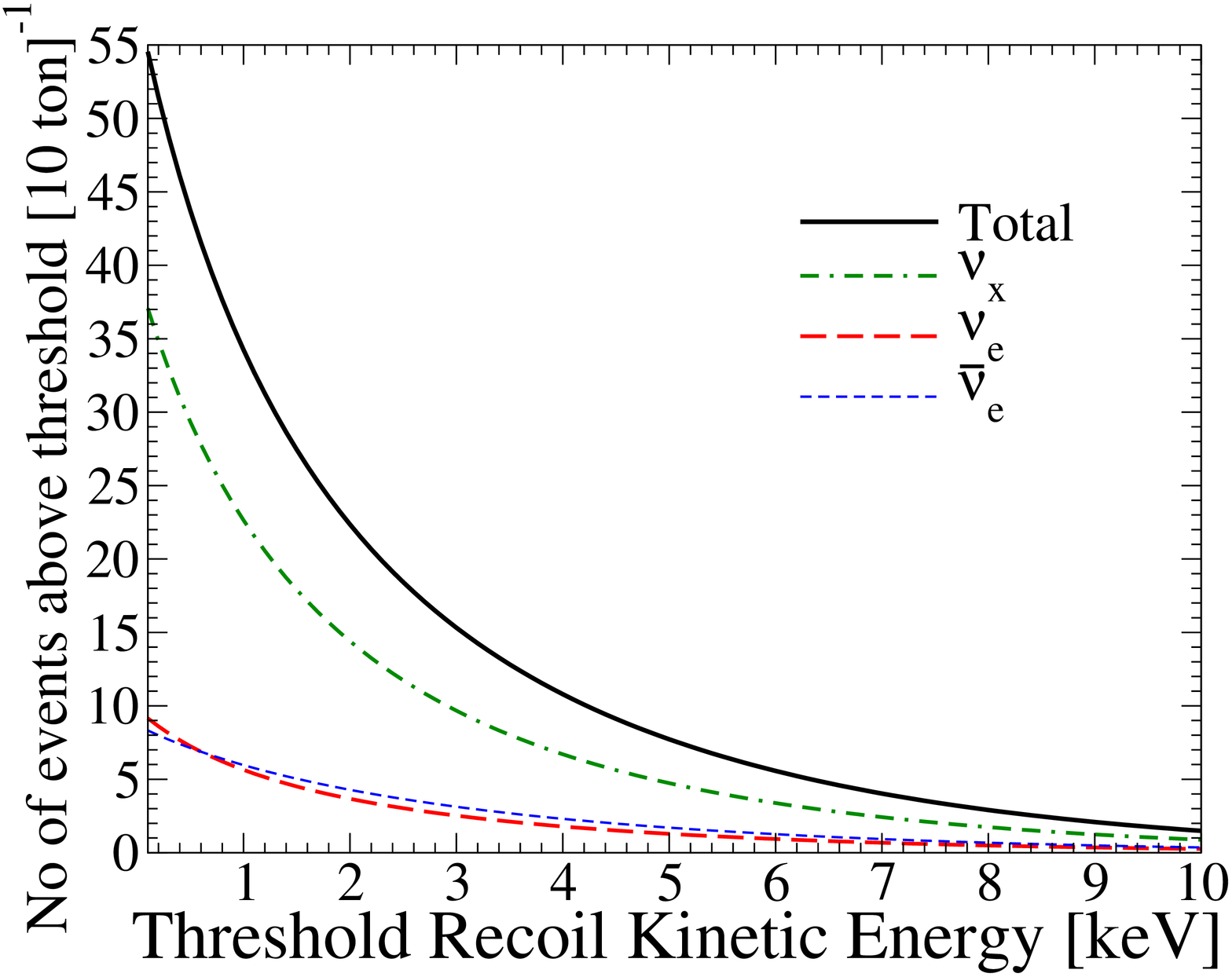,width=0.38\hsize,angle=270}
\caption{Neutrino flavor composition of the differential recoil energy 
spectra (left) and the integral recoil energy spectra above a 
threshold recoil energy as a function of the threshold recoil energy 
(right) in a 10-ton Xe detector for the accretion phase 
neutrinos from a Basel/Darmstadt SN at 10 kpc 
from earth. 
}
\label{fig:SN-Xe-nux}
\end{figure*}

\subparagraph{Extracting the $\nu_x$ properties:}
Finally, we mention an important by-product of measuring the SN 
neutrino recoil energy spectrum. For the four non-electron flavor 
neutrinos/antineutrinos, $\nu_{x}$, there are 
two physical quantities of great interest, namely, their time-averaged 
temperature and the total energy they carry. Whereas the neutral current 
reactions in conventional neutrino detectors can give information on 
their total number, measurement of the recoil spectrum for the 
CENNS events in DM detectors can in 
principle give information about the energy spectra of the neutrinos.  
Beacom, Farr and Vogel \cite{beacom-02} pointed out in connection with 
neutrino-proton elastic scattering of SN neutrinos in scintillation 
detectors like KamLAND that one can get estimates of the total 
neutrino energy as well as the temperature from the observed
proton energy spectra. Though these two observables are strongly 
correlated, a Monte Carlo simulation procedure as suggested in 
Ref.~\cite{beacom-02} may allow one to extract estimates of these two 
quantities from the observed nuclear recoil spectra in the case of 
CENNS in DM detectors. Of 
course, this is possible when the distance of the SN is known 
independently. Otherwise, like in \cite{beacom-02}, one can obtain an 
estimate of the neutrino temperature by marginalizing
over the unknown total energy. Since currently there is no 
observational handle available on the ``temperature" of the $\nu_x$ 
emitted from SNe, the possibility of measuring the SN $\nu_x$ 
temperature in DM detectors through the CENNS process is certainly worth 
exploring. 

In Figure \ref{fig:SN-Xe-nux} we plot the flavor composition of the 
expected recoil energy spectra for the accretion phase neutrinos of a 
Basel/Darmstadt SN at 10 kpc from earth in a 10 ton Xe detector. The 
main contribution to the event rate comes from $\nu_x's$ as 
they have larger average energies and also because it is a sum of 
contributions from four species of neutrinos. 

The curves shown in Figure \ref{fig:SN-Xe-nux} include  
flavor oscillation which is primarily due to MSW effect since the 
collective effects are considered matter suppressed 
\cite{Chakraborty:2011nf,Chakraborty:2011gd} during the accretion phase. 
In the cooling phase more complex scenarios of oscillations may arise
which are not very well understood \cite{Mirizzi:2011tu} and the flux expressions would be 
complicated \cite{Chakraboty:2010sz}. Hence we focus on the accretion phase
where the oscillation scenario seems settled. 
The flux differences due to different neutrino mass hierarchies are, however, not resolvable
even in the multi ton detectors. 
These curves shown in Figure \ref{fig:SN-Xe-nux} are for the 
case of inverted hierarchy. 

As already shown in Figure 
\ref{fig:nue-burst-accretion}, a 10-ton Xe detector should be able to 
distinguish the accretion phase by measuring the temporal profile of the 
neutrinos. Thus the measurement of the recoil energy spectrum together 
with the temporal profile of the events in a 10-ton Xe detector, for 
example, may indeed allow a measurement of the ``temperature" of the bulk 
of the neutrinos emitted during the main accretion phase of a SN event.     

\section{Summary and conclusion}\label{sec:summary}
In this paper we have studied the possibility of observing supernova 
neutrinos through the process of coherent elastic neutrino-nucleus 
scattering in next generation detectors for direct detection of dark 
matter. In doing this we have used the predicted neutrino flux  
from the recent Basel/Darmstadt simulations, which 
incorporate more realistic supernova physics than those used in earlier 
simulation models. We find that our estimated total event rates are 
typically a factor of 2 lower than those estimated in earlier 
studies using older simulation models. We have also studied the 
possibility of distinguishing the various phases of neutrino emission 
from the supernova through measurement of the temporal profile of the 
detected events. There we find that, with optimistic assumptions on the 
detector's time resolution ($\sim$ 10 ms) and energy threshold ($\sim$ 
0.1 keV), the neutrinos associated with the accretion 
phase of the SN at 10 kpc from earth can in principle be demarcated out 
with, for example, a 10-ton Xe detector, although distinguishing the 
neutrinos associated with the neutronization burst phase of the 
explosion would typically require several tens of ton detectors. We 
also noted that the DM detectors being flavor blind are sensitive to 
all the neutrino flavors including the non-electron flavor neutrinos, 
and thus have the potential to constrain possible non-standard physics 
of neutrino flavor oscillation. Finally, we have explored the 
possibility of measuring the temperature of the non-electron flavor 
neutrinos which are inaccessible through other charge current based detectors. 

\section*{Acknowledgments}
We are grateful to Tobias Fisher and collaborators for providing the supernova data.
We thank Rupak Mahapatra for discussions on the time resolution of the direct dark matter experiments.
One of us (PB) wishes to thank Ramanath Cowsik for hospitality and 
support under the Clark way Harrison visiting professorship  program at 
the McDonnell Center for the Space Sciences at Washington University in 
St. Louis. KK thanks Abhijit Chakrabarti for help. SC acknowledges support by a 
Marie-Curie-Fellowship of the European Community. SC also thank the Cosmology and AstroParticle division 
of Saha Institute of Nuclear Physics for kind hospitality and support during the initial phase of the project.



\end{document}